\documentclass[11pt,twoside]{article}

\usepackage{graphicx}
\usepackage{asp2006}
\usepackage{epsf}
\usepackage{psfig}
\usepackage{lscape}
\def\grtsim{\mathrel{\hbox{\rlap{\hbox{\lower2pt\hbox{$\sim$}}}\raise2pt\hbox{$>$}}}}
\def\lesssim{\mathrel{\hbox{\rlap{\hbox{\lower2pt\hbox{$\sim$}}}\raise2pt\hbox{$<$}}}}

\begin{document}
\title{Detecting galaxy clusters at 0.1 $< z <$ 2.0}   
\author{Lee Clewley, Caroline van Breukelen, David Bonfield}   
\affil{Oxford University}    

\begin{abstract} 
We present a new cluster-finding algorithm based on a combination of
the Voronoi Tessellation and Friends-Of-Friends methods. The algorithm
utilises probability distribution functions derived from a photometric
redshift analysis and is tested on simulated cluster-catalogues. We
use a 9 band photometric catalogue over 0.5 square degrees in the Subaru
XMM-Newton Deep Field. The photometry is comprised of UKIDSS Ultra
Deep Survey infrared $J$ and $K$ data combined with 3.6 $\mu$m and 4.5
$\mu$m Spitzer bands and optical $BVRi'z'$ imaging from the Subaru
Telescope. The cluster catalogue contains 13 clusters at redshifts 0.61 $\leq z \leq$
1.39 with luminosities $10 {\rm L^*} \lesssim L_{\rm tot} \lesssim 50
\rm L^*$.
\end{abstract}


\section{Introduction} 
Clusters of galaxies in the Universe play an important role in our
understanding of how dark-matter haloes collapse and large-scale
structure evolves. Their number density can place constraints on the
mass density of the Universe and the amplitude of the mass
fluctuations. Furthermore they can act as astrophysical laboratories
for understanding the formation and evolution of galaxies and their
environments. Unfortunately, there are only few clusters known at $z>1$ and
the majority of these are from X-ray surveys (e.g. using XMM-Newton;
Stanford et al. 2006). Optical searches have been stymied by the fact
that the 4000 \AA\ break falls outside the I-filter pass band ($z>1$)
given the predominance of early-type, red galaxies in
clusters. However, with the advent of large near-infrared cameras like
the Wide Field Infrared Camera (WFCAM) on the United Kingdom Infrared
Telescope (UKIRT), is is now possible to select clusters in the
near-infrared over large areas. The UKIRT Infrared Deep Sky Survey
(UKIDSS, Lawrence et al. 2006) is a suite of deep and wide surveys
using WFCAM and provides the ideal opportunity to search for
high-redshift clusters in the near infrared. The survey has a high
efficiency as WFCAM provides over an order of magnitude increase in
survey speed over previous near-infrared imagers.

There are numerous methods for detecting clusters in optical/infrared
imaging surveys. The problem is easier with spectroscopic redshifts
but large infrared spectroscopic surveys are presently impractical
over large areas. However, approximate redshifts can be calculated via
photometric redshift estimation. Despite the popularity of the photometric
redshift technique remarkably little attention is paid to using
photometric redshifts to isolate clusters. Further, little work has
been done to compare the various cluster detection method that exist
(for a review see Gal 2005).

\section{A new cluster finding algorithm}
We have developed a new cluster-detection algorithm [see van Breukelen
et al. (in prep.) for details] to deal with two common problems of
photometric selection methods: (i) projection effects of fore- and
background galaxies arising from large redshift errors ($\sigma_z \sim
0.1$); (ii) determining the reality of detected clusters arising from
cluster selection biases. We solve these issues in two ways. First, as
the photometric redshift probability functions (z-PDFs) are often
significantly non-Gaussian, and can show double peaks, our
cluster-detection algorithm samples the full z-PDF instead of a single
best redshift-estimate with an associated error. Second, selection
biases can be reduced by cross-correlating the output of two
substantially different cluster-detection methods. We therefore use
both the Voronoi Tessellation (VT) and Friends-Of-Friends (FOF)
technique to select the clusters in our survey.

In brief, our algorithm is divided into six steps: (1) We determine
the z-PDFs for all galaxies; (2) we create MC realisations of the
three-dimensional galaxy distribution, based on the galaxy z-PDFs; (3)
we divide each MC-realisation into redshift slices of $\Delta z =
0.05$ over the range $0.1 \leq z \leq 2.0$; (4) the cluster candidates
are isolated in each slice of all MC-realisations using the VT and FOF
methods; (5) the probability of cluster candidates for both methods,
from the number of MC-realisations, are mapped; (6) the output is
cross-correlated for the VT and FOF methods to produce the final
cluster-catalogue.

\subsection{Detection methods}
The VT technique works by dividing a field of galaxies into Voronoi
Cells, each containing one object: the nucleus. All points that are
closer to this nucleus than any other are enclosed by the Voronoi
Cell. One of the main advantages of this method is that it is
relatively unbiased: it does not require a particular source geometry
(e.g Ramella 2001). The parameter of interest is the area of the VT
cells, the reciprocal of which translates to a density. Overdense
regions in the plane are found by fitting a function to the density
distribution of all VT cells in the field (see Ebeling \& Wiedenmann
1993); cluster candidates are the groups of cells of a significantly
higher density than the mean background density. By contrast, the FOF
algorithm groups galaxies with a smaller separation than a projected
linking distance $D_{\rm link}$ (`friends'). A FOF algorithm utilizing
photometric redshifts was proposed by Botzler et al. (2004). A crucial
difference between our algorithm and previous work is the way we place
the galaxies in the redshift slices. We sample the full z-PDF to
create Monte-Carlo (MC)-realisations of the three-dimensional galaxy
distribution. We do not need to assign errors to individual galaxy
redshifts. An object with a large redshift error will be distributed
throughout many different slices in the MC-realisations, and therefore
not yield a significant contribution to the cluster candidates it is
potentially found in. Thus there is no need to remove objects with
large errors from the catalogue and no additional bias is introduced
against faint objects with noisier photometry. A second modification
to existing algorithms is the way we link up cluster candidates
throughout the redshift slices. Instead of comparing individual
galaxies in the clusters and linking up the clusters with
corresponding members (see e.g. Botzler et al. 2004), we use
probability maps of all redshift slices to locate likely cluster
regions.

\begin{figure}
\begin{center}
\includegraphics[height=5.5cm]{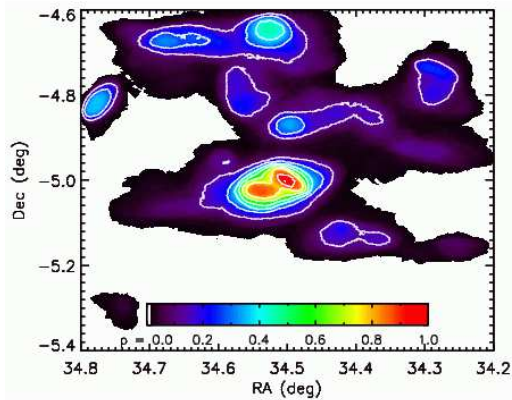}
\vspace{-3mm}
\caption{A probability map of clusters found by the Voronoi
  Tesselation method at redshift $z \sim 1.0$. Colours are normalised
  to the highest probability.}
\label{contours}
\vspace{-5mm}
\end{center}
\end{figure}

\subsection{Probability maps and cross-correlation}
We combine the cluster candidates in the redshift slices for all
MC-realisations to create a cluster probability for both methods for each
redshift slice. Figure 1 shows an example of a
probability map: the VT cluster candidates in this slice at $z = 1.0$
are contoured and coloured, with black through to red indicating low
to high probability. This map is created by overplotting the extent of
all cluster detections; the regions of the field that are found to be
in a cluster in many MC-realisations are high-probability cluster
locations. Since the error on the photometric redshifts of the
galaxies is usually larger than the width of the redshift slices, each
cluster candidate is typically found in several adjoining slices. We
join the cluster candidates that occur in the same location in several
slices by locating the peaks in the probability maps and inspecting
the area within their contours in the adjoining redshift slices for
cluster candidates. All the cluster candidates found in the same
region in adjoining redshift slices are linked up into one final
cluster; the final cluster redshift is determined by taking the mean
of the redshift slices, weighted by the number of corresponding
MC-realisations. We cross-correlate the cluster candidates output by
the VT and FOF methods and take all cluster candidates that are
reliably detected in both.

\subsection{Simulations}
To test the behaviour of our algorithm we run a set
of simulations on mock cluster catalogues ranging in total luminosity
from 10 $\rm L^*$ to 300 $\rm L^*$ and redshift $0.1 < z < 2.0$,
superimposed on a galaxy distribution randomly placed in the field
within the same redshift range. Realistic galaxy luminosities and
number densities are determined by the $K$-band luminosity function of
Cole et al. (2001) for the field-distribution and Lin et al. (2004)
for the clusters, with the simplifying assumption of passive evolution
with formation redshift $z_{form}=10$. The galaxies are spatially
distributed within a cluster according to an NFW profile (Navarro,
Frenk \& White, 1997) with a cut-off radius of 1 Mpc.

\section{Results: application to the UKIDSS-UDS cluster catalogue}
We have applied our cluster-detection algorithm to real data (van
Breukelen et al. 2006). We used: near-infrared $J$ and $K$ data from
the UKIDSS Ultra Deep Survey Early Data Release (UDS EDR, Foucaud et
al. 2006); 3.6$\mu$m and 4.5$\mu$m bands from the Spitzer Wide-area
InfraRed Extragalactic survey (Lonsdale et al. 2005); and optical
$BVRi'z'$ Subaru data over the Subaru XMM-Newton Deep Field (SXDF,
Furusawa et al. in prep.). We restricted ourselves to a rectangular
area of 0.5 square degrees, exhibiting a survey-depth of $K_{\rm
AB,lim} = J_{\rm AB,lim} = 22.5$ (UDS EDR 5$\sigma$ magnitude
limits). We included objects with a detection in $i'$, $J$ and $K$ in
the galaxy catalogue and to exclude stars and quasarswe imposed a
criterion of SExtractor stellarity index $<$ 0.8 in $i'$ and $K$. We then calculated photometric
redshifts for this sample using a modified version of \emph{hyperz}
(Bolzonella et al., 2000) which resulted in a catalogue of 19300
objects in the range $0.1 \leq z \leq 2.0$.

Application of our cluster-detection algorithm to the redshift
catalogue yielded 13 clusters at $0.61 \leq z \leq 1.39$ (van
Breukelen et al. 2006). To derive the clusters' luminosity, we
compared the results to the output of the simulations. We determined
the number of galaxies, $N_{\rm gal, FOF}$, with $K < 22.5$
(corresponding to the completeness limit) in the same way as for our
simulated clusters; this allowed us to derive an approximate total
luminosity to the cluster by interpolating between the lines of
constant total luminosity in the $N_{\rm gal}-z$ plane found in our
simulations. We found our clusters span the range of $10 {\rm L^*}
\lesssim L_{\rm tot} \lesssim 50 \rm L^*$; assuming $({M/\rm
M_{\odot}})/({L/\rm L_{\odot}}) = 75h$ (Rines et al. 2001) this yields
$0.5 \times 10^{14}~{\rm M_{\odot}} \lesssim M_{\rm cluster} \lesssim
3 \times 10^{14}~\rm M_{\odot}$.

Spectroscopic observations of these clusters are essential to confirm
their reality, particularly for the $z>1$ clusters which are
cosmologically more valuable. A new generation of highly multiplexed
near-infrared spectrometers on 8-metre class telescopes will provide
the ideal opportunity for such follow-up.
\begin{acknowledgements}
We are grateful to our other collaborators on this project, particularly the UKIDSS UDS team.
\end{acknowledgements}

\end{document}